%% file: 2019_noma_demonstrator.tex
\newcommand{\PAPERtitle}{Machine Learning-Based Adaptive Receive Filtering: Proof-of-Concept on an \acs{SDR} Platform}

\newcommand{\PAPERauthorOne}{Matthias Mehlhose}
\newcommand{\PAPERemailOne}{matthias.mehlhose}
\newcommand{\PAPERauthorTwo}{Daniyal Amir Awan}
\newcommand{\PAPERemailTwo}{daniyal.a.awan}
\newcommand{\PAPERauthorThree}{Renato L. G. Cavalcante}
\newcommand{\PAPERemailThree}{renato.cavalcante}
\newcommand{\PAPERauthorFour}{Martin Kurras}
\newcommand{\PAPERemailFour}{martin.kurras}
\newcommand{\PAPERauthorFive}{Slawomir Stanczak}
\newcommand{\PAPERemailFive}{slawomir.stanczak}

\newcommand{\PAPERemailHHI}{hhi.fraunhofer.de}
\newcommand{\PAPERemailTU}{tu-berlin.de}

\newcommand\PAPERkeywords{NOMA, 5G, machine learning, MMSE, multiuser detection}

\newcommand{\comment}[1]{}

\documentclass[conference,a4paper]{IEEEtran}

\usepackage[T1]{fontenc}
\usepackage[utf8]{inputenc}

\usepackage{cite}

\usepackage{amsmath,amssymb,amsfonts}
\usepackage{nicefrac}
\usepackage{mathtools}

\usepackage{graphicx}
\usepackage{subcaption} 

\usepackage[printonlyused,nolist]{acronym}
\usepackage[bookmarks=false]{hyperref}
\usepackage{cleveref}

\usepackage{siunitx}
\DeclareSIUnit{\sample}{Sample}
\DeclareSIQualifier\isotropic{i} 
\DeclareSIQualifier\milli{m} 

\usepackage{marginnote}
 
\AtBeginDocument{\title{\PAPERtitle}}
\AtBeginDocument{
    \author{
        
        \IEEEauthorblockN{\PAPERauthorOne\IEEEauthorrefmark{1}, \PAPERauthorTwo\IEEEauthorrefmark{2}, \PAPERauthorThree\IEEEauthorrefmark{1}\IEEEauthorrefmark{2}, \PAPERauthorFour\IEEEauthorrefmark{1}\ and \PAPERauthorFive\IEEEauthorrefmark{1}\IEEEauthorrefmark{2}}
        
       \vspace{0.05in}
        \IEEEauthorblockA{\IEEEauthorrefmark{1}
            \textit{Fraunhofer Institute for Telecommunications},
            \textit{Heinrich Hertz Institute (HHI)}\\
            Department of Wireless Communications and Networks,
            Einsteinufer 37, 10587 Berlin, Germany\\
            \{\PAPERemailOne, \PAPERemailThree, \PAPERemailFour, \PAPERemailFive \}@\PAPERemailHHI
        }
        
        \vspace{0.05in}
        \IEEEauthorblockA{\IEEEauthorrefmark{2}
            \textit{Technical University Of Berlin},
            \textit{Faculty IV - Electrical Engineering and Computer Science}\\
            Department of Telecommunication Systems,
            Einsteinufer 27, 10587 Berlin, Germany\\
            \{\PAPERemailTwo, \PAPERemailThree, \PAPERemailFive \}@\PAPERemailTU
        }
        
        \vspace{-0.10in}
        
    }
}

\AtEndDocument{\bibliographystyle{IEEEtran}}
\AtEndDocument{\bibliography{IEEEabrv,bib/noma_demonstrator_book,bib/noma_demonstrator_paper,bib/noma_demonstrator_techrep,bib/noma_demonstrator_url,bib/noma_demonstrator_diss}}

\usepackage{tikz}
\usetikzlibrary{shapes,arrows}

\graphicspath{{./pictures/}{../matlab/data/}}

\hypersetup{
  pdftitle    = {\PAPERtitle},
  pdfauthor   = {\PAPERauthorOne, \PAPERauthorTwo, \PAPERauthorThree, \PAPERauthorFour\ and \PAPERauthorFive},
  pdfkeywords = {\PAPERkeywords} ,
  pdfsubject  = {Conference paper},
  pdfcreator  = {XeLatex},
  pdfproducer = {LaTeX with hyperref},
  pdflang={english},
}

\hypersetup{
}

\hypersetup{
	colorlinks = true,
	linkcolor  = {black},
	citecolor  = {black},
	urlcolor   = {black},
}

\hypersetup{
	bookmarks=false,
}

\begin{document}

\renewcommand{\IEEEiedlistdecl}{\IEEEsetlabelwidth{xxxxx}}
\begin{acronym}

    \acro{4G}[4G]{4th Generation}
    \acro{5G}[5G]{5th Generation}
    \acro{5GPPP}[5G-PPP]{5G Infrastructure Public Private Partnership}
    \acro{3GPP}[3GPP]{3rd Generation Partnership Project}
    \acro{ADC}[ADC]{analog-to-digital converter}
    \acro{APSM}[APSM]{adaptive projection subgradient method}
    \acro{BER}[BER]{bit error rate}
    \acro{BPSK}[BPSK]{binary phase-shift keying}
    \acro{BS}[BS]{base station}
    \acro{CFO}[CFO]{carrier frequency offset}
    \acro{COMP}[CoMP]{coordinated multi-point}
    \acro{COTS}[COTS]{commercial off-the-shelf}
    \acro{CPU}[CPU]{central processing unit}
    \acro{DAC}[DAC]{digital-to-analog converter}
    \acro{DL}[DL]{downlink}
    \acro{DMA}[DMA]{direct memory access}
    \acro{EC}[EC]{European commission}
    \acro{eNB}[eNB]{Evolved Node B}
    \acro{FPGA}[FPGA]{field-programmable gate array}
    \acro{FR1}[FR1]{Frequency Range 1}
    \acro{FR2}[FR2]{Frequency Range 2}
    \acro{gNB}[gNB]{next generation Node B} 
    \acro{H2020}[H2020]{Horizon 2020}
    \acro{IIO}[IIO]{industrial I/O}
    \acro{IOT}[IoT]{internet of things}
    \acro{LO}[LO]{local oscillator}
    \acro{MAP}[MAP]{maximum a-posteriori}
    \acro{MCM}[MCM]{multi-carrier modulation}
    \acro{ML}[ML]{Machine Learning}
    \acro{MIMO}[MIMO]{multiple-input multiple-output}
    \acro{MMIMO}[mMIMO]{Massive Multiple-Input Multiple-Output}
    \acro{MMSE}[MMSE]{minimum mean square error}
    \acro{MMTC}[mMTC]{massive machine type communications}
    \acro{NOMA}[NOMA]{non-orthogonal multiple access}
    \acro{MUSA}[MUSA]{multi-user Shared Access}
    \acro{NL}[NL]{Non-Linear}
    \acro{NR}[NR]{New Radio}
    \acro{OFDMA}[OFDMA]{orthogonal frequency-division multiple access}
    \acro{OMA}[OMA]{orthogonal multiple access}
    \acro{PC}[PC]{personal computer}
    \acro{PLL}[PLL]{phase lock loop}
    \acro{RKHS}[RKHS]{reproducing kernel Hilbert space}
    \acro{RRH}[RRH]{remote radio head}
    \acro{SCM}[SCM]{Single-Carrier Modulation}
    \acro{SC-FDMA}[SC-FDMA]{Single-Carrier Frequency-Division Multiple Access}
    \acro{SDR}[SDR]{Software-Defined Radio}
    \acro{SIC}[SIC]{successive interference cancellation}
    \acro{SIMO}[SIMO]{Single-Input Multiple-Output}
    \acro{SINR}[SINR]{signal-to-noise-plus-interference-ratio}
    \acro{SNR}[SNR]{signal-to-noise-ratio}
    \acro{SoC}[SoC]{system on chip}
    \acro{SER}[SER]{symbol error rate}
    \acro{TS}[TS]{technical specification}
    \acro{UE}[UE]{user equipment}
    \acro{UL}[UL]{uplink}
    \acro{URLLC}[URLLC]{ultra-reliable and low latency communications}
    \acro{UPA}[UPA]{uniform planar antenna}
    \acro{WLAN}[WLAN]{Wireless Local Area Network}
	
\end{acronym}
\renewcommand{\IEEEiedlistdecl}{\relax}


\maketitle

\begin{abstract}%
\input{sections/abstract.tex}

\end{abstract}

\smallskip 
\begin{IEEEkeywords}
\PAPERkeywords
\end{IEEEkeywords}


 
\input{sections/introduction}
\input{sections/materials_and_methods}
\input{sections/results}

\input{sections/conclusions}
\input{sections/acknowledgments}
\vfill

\end{document}

%% file: sections/abstract.tex
\relax
\relax   
\relax
\relax
Conventional multiuser detection techniques either require a large number of antennas at the receiver for a desired performance, or they are too complex for practical implementation.
Moreover, many of these techniques, such as \ac{SIC}, suffer from errors in parameter estimation (user channels, covariance matrix, noise variance, etc.) that is performed before detection of user data symbols. As an alternative to conventional methods, this paper proposes and demonstrates a low-complexity practical \ac{ML} based receiver that achieves similar (and at times better) performance to the \ac{SIC} receiver. The proposed receiver does not require parameter estimation; instead it uses supervised learning to detect the user modulation symbols directly.
We perform comparisons with \ac{MMSE} and \ac{SIC} receivers in terms of \ac{SER} and complexity.


\comment{
\marginnote{\textcolor{red}{Reason for this investigation}}
\ac{NOMA} functions enable upcoming generation mobile communication systems like \ac{5G} \ac{NR} to make better use of time and frequency resources by sharing them overall or a subset of users.
The ability to share resources in the \ac{UL} of \ac{NR} received a lot of research attention and was investigated in the \ac{3GPP} Study Item \cite {3GPP-TR-38.812}.

\marginnote{\textcolor{red}{Objective of this investigation}}
In combination with the \ac{MIMO} technology, the classification and separation of the transmitter signal of one specific user or user group from noise and the interference of the other users by beamforming become feasible.
Therefore, this \ac{NOMA} experiment focuses on users that are distinguishable in spatial or in power domain or both. Multiplexing of users in code-domain is not used in this demonstration. 
The \ac{RKHS} based machine learning algorithm used in this paper is based on the combination of the weighted sum of a linear and a non-linear filter.
Because linear receive beamforming techniques is robust, but cannot deal with the resulting excessive multiple-access interference adequately.
Especially for the case that we have more transmit antennas active than antennas available on the receiver side.

\marginnote{\textcolor{red}{What was investigated}}
In this paper, we examine the implementation complexity and performance of our \ac{ML} based adaptive \ac{NOMA} uplink scheme compared to a standard linear algorithm like \ac{MMSE} and non-linear algorithm like \ac{MMSE}-\ac{SIC}.
We have investigated our \ac{NOMA} approach in the sub \SI{6}{\giga\hertz} range.

\marginnote{\textcolor{red}{Results of this investigation}}
Our demo system shows a modular and compact prototype approach for performance measurements of a \ac{MMIMO} \ac{NOMA} system.
The experimental demonstration has shown that \ac{NOMA} can meet both, the performance in the region of non-linear algorithms like \ac{MMSE}-\ac{SIC} with a detection delay in the region of linear algorithms like \ac{MMSE}.

\marginnote{\textcolor{red}{Importance of the investigation}}
We show the possibility that each user can be detected independently and therefore in parallel with other users.
The resulting fact, that the detection delay of our \ac{ML} \ac{NOMA} approach is low leads to a small latency of the overall system and thus meets one of the major requirements of \ac{NR}.

}

%% file: sections/introduction.tex

\section{Introduction}\label{sec:Intro}
To fulfill the requirement of \textit{massive connectivity} in \ac{5G} mobile networks, highly efficient and flexible use of time and frequency resources is necessary.
Current mobile networks mostly employ \ac{OMA} radio access techniques such as \ac{OFDMA}.
In these systems, at each \ac{BS}, a scheduler allocates resources to each user so as to avoid strong interference from other users. 
However, the \textit{massive connectivity} requirement in \ac{5G} means that the number of simultaneously transmitting devices in the uplink is expected to exceed the number of available orthogonal resources. Therefore, recently a significant body of research has proposed \ac{NOMA} \cite{3GPP-TR-38.812,5256321,7174566,8464916,8417843,7973146,7842464,6692652,DBLP:journals/corr/abs-1801-05541}. 
In these systems, users are allowed to share resources (and thus interfere) and \textit{multiuser detection} techniques \cite{Verdu} are employed to perform reliable detection of user symbols in the uplink. It is well-known that, though the optimal multiuser receiver (in terms of \ac{BER}) is nonlinear \cite{Verdu}, if the number of antennas at the BS receiver exceeds the number of users, then low-complexity linear receivers (e.g., the \ac{MMSE} receiver) often show a good performance. However, in \textit{massive connectivity} scenarios, a very large number of antennas at the receiver are required to achieve a good performance using linear techniques. Furthermore, it is well known that massive antenna systems suffer from various problems such as pilot contamination \cite{7339665}.
Therefore, the assumption that the number of antennas at the \ac{BS} will be larger than the number of users may not be realistic. In this case, nonlinear detection methods have to be considered.
In this study, we demonstrate a promising machine learning based nonlinear multiuser receiver (see \cite{8422449}) that does not require explicit knowledge of desired or interfering user channels or powers. Furthermore, all users are detected in parallel and independently, and the technique has a relatively low complexity.

\subsection{The Motivation for a Machine Learning Receiver}\label{sec:motivation_for_a_machine_learning_receiver)}
As mentioned above, it is known that the optimal detector is the (nonlinear) \ac{MAP} detector \cite{Mitra1994}. Unfortunately, the \ac{MAP} detector requires exact knowledge of user channels, their transmit powers and variance of system noise. Estimation of these parameters adds considerable complexity to the receiver design and their estimation is subject to errors.
As a compromise, suboptimal nonlinear receivers such as the \ac{SIC} receiver and those based on neural networks have been proposed \cite{Isik2007,Mitra1994,Aazhang1992}.
However, these receivers also require a good estimation of the above mentioned parameters and their complexity becomes impractical for a large number of users.
Due to these issues with conventional receiver design, \cite{8422449} proposed a \ac{ML} based technique that simplifies receiver design by using supervised learning to detect the desired user symbols directly without any intermediate parameter estimation.
It has been shown in \cite{8422449} through simulations that this technique outperforms the \ac{MMSE}-\ac{SIC} receiver (even in the case of exact parameter knowledge) when the number of users exceeds the number of antennas at the \ac{BS}. Furthermore, this technique works with a relatively small training set which is an important requirement (especially, in the physical layer) in dynamic wireless environments. In the following, we present the multiuser receive filtering model of \cite{8422449} that we implement and demonstrate in the study.

\subsection{Multiuser Receive Filtering Model}\label{sec:multiuser_receive_filtering_model}
In the uplink setup, $K \in \mathbb{N}$ symbol-synchronized users transmit to the \ac{BS} equipped with $M\in \mathbb{N}$ antennas at the same frequency.
Assuming a non-dispersive channel and symbol-time sampling at time $t \in \mathbb{N}$, we can write the received baseband signal as
\begin{equation}
\label{eqn:systemmodel}
 \mathbf{r}\left(t\right) = \sum_{k=1}^{K}    \sqrt{p_k\left(t\right)}    b_k\left(t\right)    \mathbf{h}_k\left(t\right)    +\mathbf{n}\left(t\right)    \in \mathbb{C}^{M}
\end{equation}
where $\mathbf{h}_k\left(t\right) \in \mathbb{C}^{M}$, $b_k\left(t\right)  \in \mathbb{C}^{M}$, and $p_k\left(t\right)  \in \mathbb{C}^{M}$ is the channel, the information-bearing symbol, and the transmit power, respectively, of the $k$th user, and where $\mathbf{n}\left(t\right) \in \mathbb{C}^{M}$ denotes additive noise plus interference from unknown transmissions.
The job of baseband receive-filtering is to detect the modulation symbol $b_k(t)$ of each desired user given $\mathbf{r}\left(t\right)$.

The online adaptive filtering algorithm of \cite{8422449} performs the receive filtering for each desired user independently in parallel.
In more detail, data communication is preceded by a training phase in which desired users transmit sequences $(\forall k \in \overline{1,K})$ $(b_k)_{t \in \overline{1,T_{\text{length}}}}$, where $T_{\text{length}}$ denotes training time, to the \ac{BS}.
Suppose we are detecting user $k=1$, then the algorithm has access to the training data $(b_1(t),\mathbf{r}\left(t\right))_{t \in \overline{1,T_{\text{length}}}}$. The training is performed by an \ac{APSM} \cite{Yamada} based algorithm.
Since this algorithm operates in a \ac{RKHS} $\mathcal{H}_{\text{sum}}$, consisting of weighted linear and nonlinear functions, most operations required by \ac{APSM} training are carried out via inner products.
Therefore, the algorithm has a low complexity. During training $T_{\text{length}}$, and for the $k$th user, the algorithm learns a nonlinear filter $f \in \mathcal{H}_{\text{sum}}$ such that
\begin{equation}
(\forall t \in \mathbb{N})~|f(\mathbf{r}(t))-b_k(t)|\leq \epsilon,
\label{eqn:convex_set} 
\end{equation}
where $\epsilon>0$ is a sufficiently small (typically $\epsilon=0.01-0.1$) design parameter tuned such that correct hard-detection of $b_k(t)$ is possible. For sufficiently large training symbols $T_{\text{length}}$, and under certain assumptions \cite{6494588,Yamada}, the algorithm obtains a good estimate of an $f^{\ast} \in \mathcal{H}_{\text{sum}}$ satisfying \eqref{eqn:convex_set}. For more details, see \cite{8422449}.    

\subsection{Contribution and Paper Structure}
We summarize the contributions of this study in the following:
\begin{enumerate}
\item In our \textit{hardware-in-the-loop} setup, we demonstrate our proposed \ac{ML} based multiuser receiver that is a simpler yet powerful alternative to conventional multiuser receivers. 
\item We compare the performance of our receiver with the linear \ac{MMSE} and the nonlinear \ac{MMSE}-\ac{SIC} receivers. We show that our receiver outperforms \ac{MMSE} and it shows comparable (and at times better) performance to \ac{MMSE}-\ac{SIC}. Furthermore, we demonstrate that this receiver has a low complexity.   
\end{enumerate}  

The remainder of the study is organized as follows.
In Section \ref{sec:MatMet}, we present our receiver design including the hardware equipment that we employ during the demonstration and the signal processing.
In Section \ref{sec:Res}, we present our results and perform comparisons with conventional techniques.
Section \ref{sec:Con} concludes this study.

\comment{
\marginnote{\textcolor{red}{Extended reason for this research}}
New major functionalities for different use cases are added to the system with each new mobile network generation.
To fulfill the requirements in \ac{5G}, very flexible use of time and frequency resource and new case families like massive \ac{IOT}, \ac{URLLC} and \ac{MMTC} are added in particular \cite{5GPPP_WP}.
One goal of each new generation was, that the number of simultaneously operating devices within the coverage of each cell also increases.
To archive this in a \ac{4G} network, the number of antennas at the base station had to increase as well to enable \ac{MIMO} and \ac{COMP} operations.
Although the available space for the antennas can be better exploited by smaller antennas at higher frequencies.
Nevertheless, the maximum number of antennas is still limited depending on the location.
Also the distances between the antennas of different cells cannot be reduced arbitrarily.
The results of the regular auctions of different frequency bands make it clear that bandwidth is a rare and very expensive resource.
For the classical approach of simply increasing the bandwidth, blocks with a total bandwidth in the GHz range were reserved in \ac{FR2} from \SIrange[range-units=single,range-phrase=\ to\ ]{24250}{52600}{\mega\hertz}, compared to \ac{FR1} from \SIrange[range-units=single,range-phrase=\ to\ ]{410}{7125}{\mega\hertz} \cite{3GPP-TS-38.104}.
A promising approach is the combination of non-linear algorithms in combination with machine learning tools.
In contrast to classical \ac{SIC} algorithms, these offer the possibility to detect the superimposed (non-orthogonal) user signals at the base station in parallel, i.e. simultaneously, without dependencies and knowledge of the other user signals.
When the \ac{gNB} is equipped with an array antenna, an \ac{MMSE} beamforming algorithm can also be taken into account.
}

\comment{
The contribution of this study is the demonstration of a new ML-based multiuser demodulation technique in the laboratory. We demonstrate the potential and advantage of machine learning   

show how the novel machine learning based filter \cite{8422449} is performing with state of the art hardware and also under a specific environment like indoor \ac{MIMO} channel and limited bandwidth.
The performance as well as the processing delay will  be compared to \ac{MMSE} and \ac{MMSE}-\ac{SIC}.
Hardware impairments like different phases at the antenna ports are not calibrated, they are introduced by different track length on the \ac{PCB} as well as cable length and adapters between \ac{SDR} and antennas.
In addition, they are multiple \acp{PLL} and clock buffers in the clock distribution network on each \ac{SDR} module, which can also introduce random phases at each reset.
Both, receiver as well as transmitter \ac{SDR} are able to share the same reference clock to keep the residual \ac{CFO} low in this setup.
Several active users are arranged as one single cluster transmitting well know \ac{BPSK} signals in the time domain.
}

%% file: sections/materials_and_methods.tex

\section{Experiment Setup}
\label{sec:MatMet}
This section provides the technical details of the hardware and software components being used for our \ac{NOMA} enabled \ac{SDR} \textit{hardware-in-the-loop} setup.

\begin{figure}[htbp]
    \captionsetup{skip=0.5\baselineskip,size=footnotesize}
    \centering
    \includegraphics[width=0.97\linewidth,trim={1.0cm 0.0cm 1.5cm 1.0cm},clip]{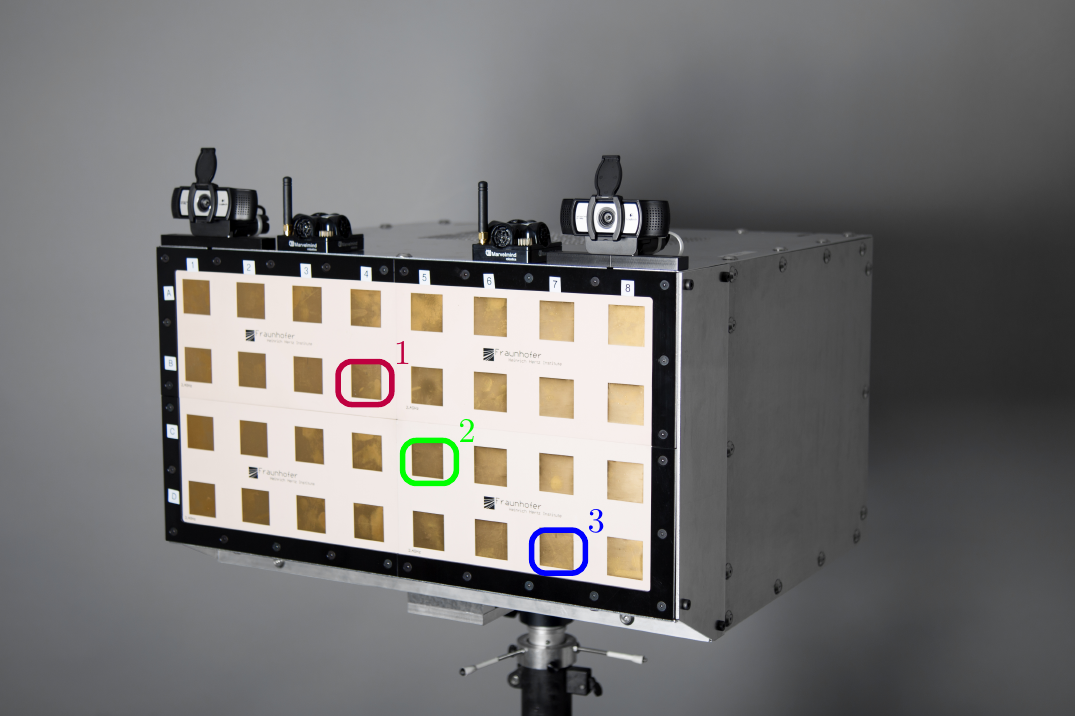} 
    \caption[BSA]{\acs{BS} \acs{WLAN} \acs{UPA} Array}
    \label{fig:UPA}
    \vspace*{0.5\floatsep}
    \captionsetup{skip=0.5\baselineskip,size=footnotesize}
    \centering
    \includegraphics[width=0.97\linewidth,trim={4.0cm 0.0cm 1.0cm 2.5cm},clip]{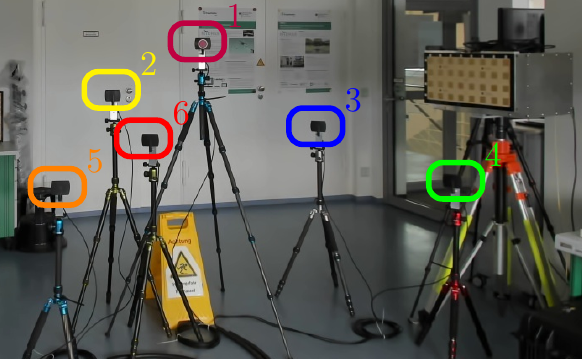} 
    \caption[BSA]{Demo Setup}
    \label{fig:LAB_UE}
    \vspace*{-1.0\floatsep}
\end{figure}

\subsection{Equipment}

\subsubsection{\acl{SDR}}
On the digital side of our \ac{SDR} a Xilinx Zynq XC7Z045 \ac{SoC} module is mounted.
This \ac{SoC} combines both a Linux running dual ARM Cortex-A9 core pack and \ac{FPGA}, well equipped to handle both the control and the signal processing tasks.
On the analog side, four independent Analog Devices AD9361 are used, each containing an integrated two transceiver path 12-bit \ac{ADC} and \ac{DAC} solution.
Each \ac{SDR} module \cite{NAMC-SDR} has eight antenna connections for the transmission and reception direction, respectively. 
All \ac{SDR} modules are able to share a \SI{10}{\mega\hertz} reference clock.

\subsubsection{Base Station}
The array antenna, shown in \Cref{fig:UPA}, is equipped with 32 equidistant cross-polarized patch antenna elements arranged in 4 rows and 8 columns on a planar printed circuit board.
Each patch element of this array operates at a center frequency of \SI{2.442}{\giga\hertz}, which correspond to \ac{WLAN} channel 7.
The element spacing is $\nicefrac{\lambda}{2}$, where $\lambda = \SI{12.28}{\centi\meter}$ refers to the wavelength at resonant frequency $f = \SI{2.442}{\giga\hertz}$, and the length and width of each quadratic patch element is $\nicefrac{\lambda}{4}$.
Each antenna housing has space for up to four \ac{SDR} modules. 
The \ac{BS} parameters are summarized in \cref{tab:BSparams}.

\begin{table}[htbp]
\caption{\ac{BS} Parameters}
\setlength\tabcolsep{2pt}
\begin{center}
\begin{tabular}{|c|c|c|}
\hline
parameter & value & comment \\
\hline
resonant frequency & \SI{2.442}{\giga\hertz} & \\
\hline
element spacing & \SI{6.14}{\centi\metre} & $\nicefrac{\lambda}{2}$ at \SI{2.442}{\giga\hertz} \\
\hline
element length and width & \SI{3.07}{\centi\metre} & $\nicefrac{\lambda}{4}$ at \SI{2.442}{\giga\hertz} \\
\hline
number of rows & 4 & \\
\hline
number of columns & 8 & \\
\hline
polarization & H and V & vertical is used \\
\hline
number of \acs{SDR} modules & up to 4 & \\
\hline
number of module antenna ports & 8  & 4x AD9361 \\
\hline
\end{tabular}
\label{tab:BSparams}
\end{center}
    \vspace*{-1\floatsep}
\end{table}
%
\subsubsection{User Equipment}
Each of the six \acp{UE}, shown in \Cref{fig:LAB_UE}, is equipped with a \ac{COTS} \ac{WLAN} antenna which is connected to one of eight \ac{SDR} module transmitter ports. 
According to the datasheet of the manufacturer, the measured antenna gain is \SI[qualifier-mode = text]{5}{\deci\bel\isotropic} at  \SI{2.5}{\giga\hertz} and up to \SI[qualifier-mode = text]{7}{\deci\bel\isotropic} at  \SI{5.7}{\giga\hertz} and the polarization is vertical.


%
\subsection{Signal Processing}
\begin{figure*}[htbp]
    \captionsetup{skip=0.5\baselineskip,size=footnotesize}
    \centering
    \includegraphics[width=0.97\linewidth]{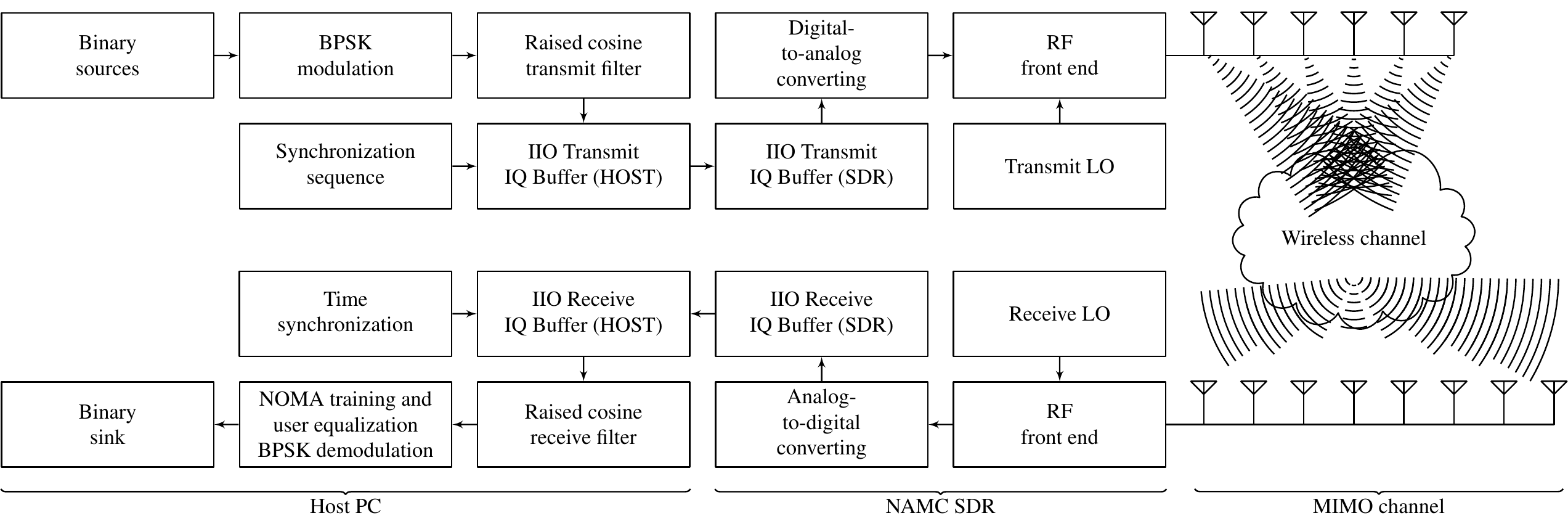}
    \caption[SM]{System Model}
    \label{fig:SM}
    \vspace*{-1.0\floatsep}
\end{figure*}

In the demonstration, we use an \ac{SDR} based transceiver connected to a subset of patch antennas arranged as a \ac{UPA} array with four rows and eight columns at the receive side.
On the transmit side we use a single \ac{SDR} connected to all users. Each transmit port is connected to a single-antenna user.
\Cref{tab:SYSparams} sums up the demonstration system parameters.
The raw time data IQ-buffers from each \ac{SDR} are transmitted and received over a \SI{1}{\giga\bit} Ethernet interface on each \ac{SDR} module. 
Transmit signal creation for the users and received signal processing is done on a standard desktop \ac{PC}. 
The system architecture of the transmit and receive path is shown in \Cref{fig:SM}. 

\subsubsection{Transmit Signal Chain}
On the transmit side \ac{BPSK} symbols at each user are filtered by a raised cosine transmit filter with an oversampling factor of $16$ to avoid excessive multipath transmissions. The training sequence $(b_k)_{t \in \overline{1,T_{\text{length}}}}$ and the data sequence $(b_k)_{t > T_{\text{length}}}$ (see Section \ref{sec:multiuser_receive_filtering_model}) are separated by a \textit{Frank-Zadoff-Chu} sequence that is used for time synchronization on the receive side. 

\subsubsection{Receive Signal Chain}
On the receive side, we receive an IQ buffer twice the size of the transmit buffer size.
During each run of the \textit{hardware-in-the-loop} experiment, the filled receive buffer contains a full radio frame of training, synchronization, and data sequences and parts of the preceding and following radio frames.
We perform a cross-correlation based time-synchronization to find the position of a complete frame and extract the superimposed received signals corresponding to the training and data sequences from each user. Both sequences are then filtered with a raised cosine receive filter and sub-sampled by a factor of $16$.
The first part of this filter output is used to train the \ac{ML} algorithm and the second part is used to detect the \ac{BPSK} symbols.
Finally, we compare the bit sequence in the binary sink with the bit sequence from the binary source to calculate the \ac{SER} for each user. Note that we can only measure the SER over the length of the SDR buffer.

\begin{table}[htbp]
\caption{Proof-of-Concept System Parameters}
\setlength\tabcolsep{2pt}
\begin{center}
\begin{tabular}{|c|c|c|}
\hline
parameter & value & comment \\
\hline
carrier frequency & \SI{2.442}{\giga\hertz} & WiFi Band 7 (tune able)\\
\hline
system bandwidth & \SI{30.72}{\mega\hertz} & \\
\hline
\ac{NOMA} bandwidth & \SI{1.92}{\mega\hertz} & \footnotesize{oversampling by a factor of 16} \\
\hline
\ac{NOMA} waveform & \acs{BPSK} & \acf{SCM} \\
\hline
\ac{BS} height & \( \approx \SI{1.5}{\meter} \) & \\
\hline
number of rx antennas & \( 3 \) & see \Cref{fig:UPA} \\
\hline
number of tx antennas & \( \leq 6 \) & see \cref{fig:LAB_UE} (active users) \\
\hline
user transmit power & \( \SIrange[qualifier-mode = text,range-units=single,range-phrase=\ to\ ]{-15}{0}{\deci\bel\milli} \) & difference of \SI{3}{\deci\bel} between users\\
\hline
user height distribution & \( \SIrange[range-units=single,range-phrase=\ to\ ]{0.8}{1.8}{\meter} \) & \\
\hline
distance \ac{BS} and users & \( \SIrange[range-units=single,range-phrase=\ to\ ]{2.6}{4.0}{\meter} \) & \\
\hline
training symbols & \( \leq 685 \) & default: 500 \\
\hline
data symbols & \( \leq 3840 \) & default: 3000 \\
\hline
\end{tabular}
\label{tab:SYSparams}
\end{center}
    \vspace*{-1\floatsep}
\end{table}

%% file: sections/results.tex

\section{Demonstration Results}
\label{sec:Res}

In this section, we present the performance of our \textit{hardware-in-the-loop} demonstrator. 
We compare the performance of the \ac{ML}-based multiuser receive filtering (see Section \ref{sec:multiuser_receive_filtering_model}), which we denote by \ac{NL}-\ac{ML} in the following, with:
\begin{itemize}
\item the linear \ac{MMSE} receive filtering in which each user is detected in parallel, and
\item the nonlinear \ac{MMSE}-\ac{SIC} receiver. In \ac{MMSE}-\ac{SIC}, each user is first detected using \ac{MMSE} filtering and then its contribution is subtracted from the received signal \eqref{eqn:systemmodel} before the next user in the \ac{SIC}-chain is detected. 
\end{itemize}

\subsection{Demonstration Setting}

As mentioned in Section \ref{sec:multiuser_receive_filtering_model} the data communication is preceded by a training phase of length $T_{\text{length}}$ complex signals.
This training sequence is used to train the \ac{NL}-\ac{ML} algorithm and also to perform channel and covariance matrix estimation for \ac{MMSE} filtering.
We use Algorithm 1 in \cite{8422449} to perform training for each user in parallel. The important parameters are shown in \Cref{tab:noma_parameters} and other parameters were chosen as in \cite{8422449}.
There are $K=6$ desired users in the system and we use $M=3$ antennas at the BS. 
In order to multiplex the users in the power domain, as required in power-domain \ac{NOMA}, each successive user transmits with \SI{3}{\decibel} less power.


\begin{table}[htbp]
\caption{\ac{NL}-\ac{ML} Parameters}
\setlength\tabcolsep{10pt}
\begin{center}
\begin{tabular}{|c|c|c|c|c|}
\hline
User & $w_{\text{L}}$ & $w_{\text{G}}$ &  $T_{\text{length}}$ & $D_{\text{length}}$ \\
\hline
1 & 60\% & 40\% & 10 & 3000 \\
\hline
2 & 60\% & 40\% & 70 & 3000 \\
\hline
3 & 60\% & 40\% & 40 & 3000 \\
\hline
4 & 60\% & 40\% & 40 & 3000 \\
\hline
5 & 60\% & 40\% & 685 & 3000 \\
\hline
6 & 60\% & 40\% & 685 & 3000 \\
\hline
\end{tabular}
\label{tab:noma_parameters}
\end{center}
    \vspace*{-1\floatsep}
\end{table}

\subsection{Comparison with \ac{MMSE} and \ac{MMSE}-\ac{SIC}}
In this section we compare the SER and complexity performance of  \ac{NL}-\ac{ML} with that of the \ac{MMSE} and \ac{MMSE}-\ac{SIC}. As mentioned above, users are separated in the power domain where user $1$ is the ``strongest'' user and user $6$ is the ``weakest'' user in the system. In terms of the \ac{SER}, we observed that all $3$ techniques showed a comparable near-perfect performance for the first $4$ users, i.e., $\text{\ac{SER}}\approx 0$ over the observable test sample set (which was restricted to a size of $3000$ symbols). Therefore, we omit the results for these users. 

In the following, we show the performance for the $5$th and the $6$th users (which can be thought of as weak users) in terms of the SER and complexity of detection (that we measure in terms of the processing delay). Note that, due to the limited size of the \ac{SDR} buffer, we can only observe the detection errors for $3000$ symbols. Therefore, the minimum observable non-zero SER is $3.33 \times 10^{-4}$. Additionally, we show the performance of the overall system in terms of the average SER and average complexity/delay.  

\begin{figure}[b!]
   \vspace*{-1.0\floatsep}
    \captionsetup{skip=0.2\baselineskip,size=footnotesize}
    \centering
    \includegraphics[width=0.97\linewidth,clip=true]{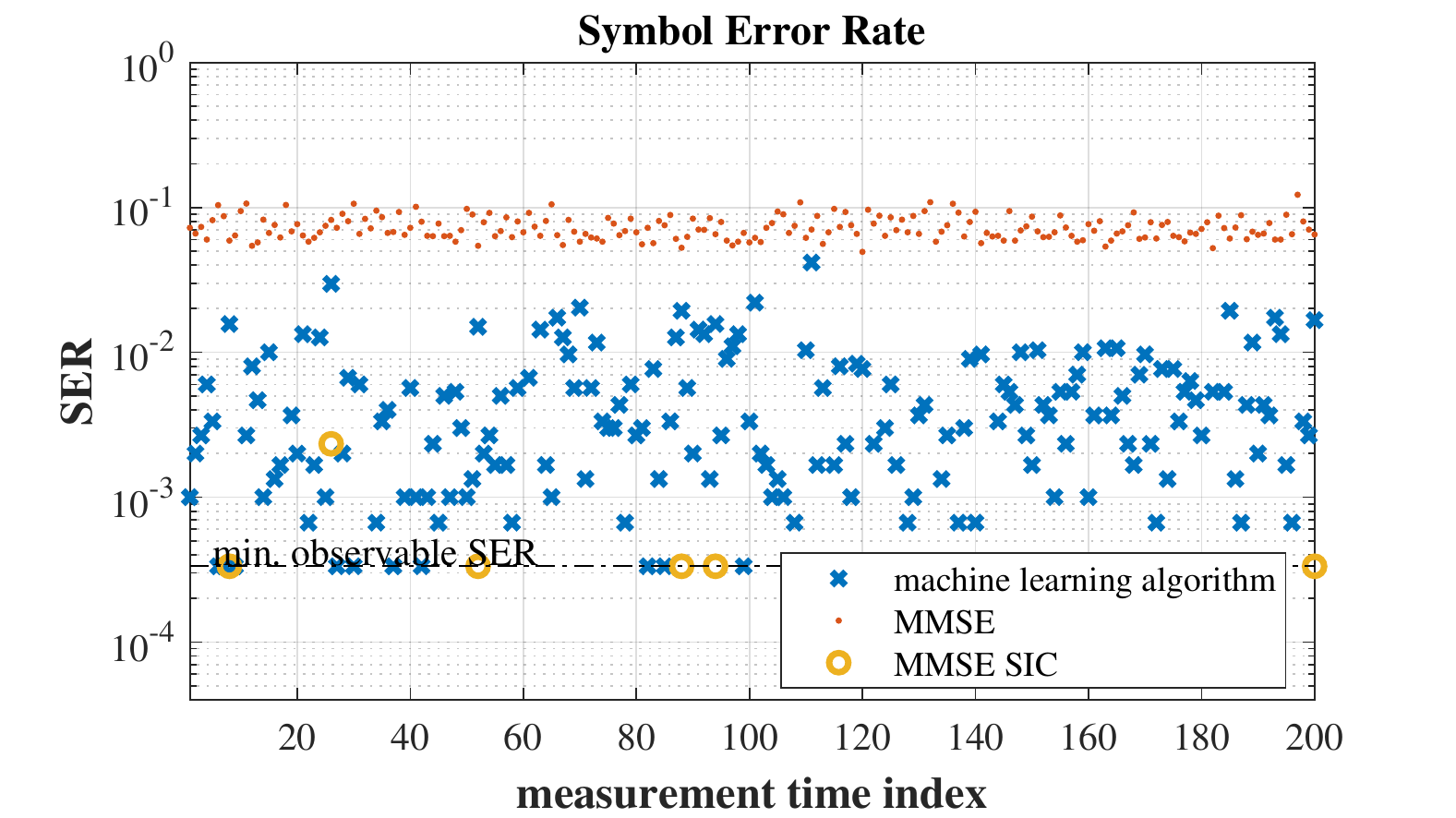}
    \caption[SER5]{User 5: \(  \omega_{L} = 0.6 \) and \( \omega_{G} = 0.4 \) with \( T_{length} = 685 \)}
    \label{fig:SER5}
    \vspace*{+0.5\floatsep}
    %
%
    \captionsetup{skip=0.2\baselineskip,size=footnotesize}
    \centering
    \includegraphics[width=0.97\linewidth,clip=true]{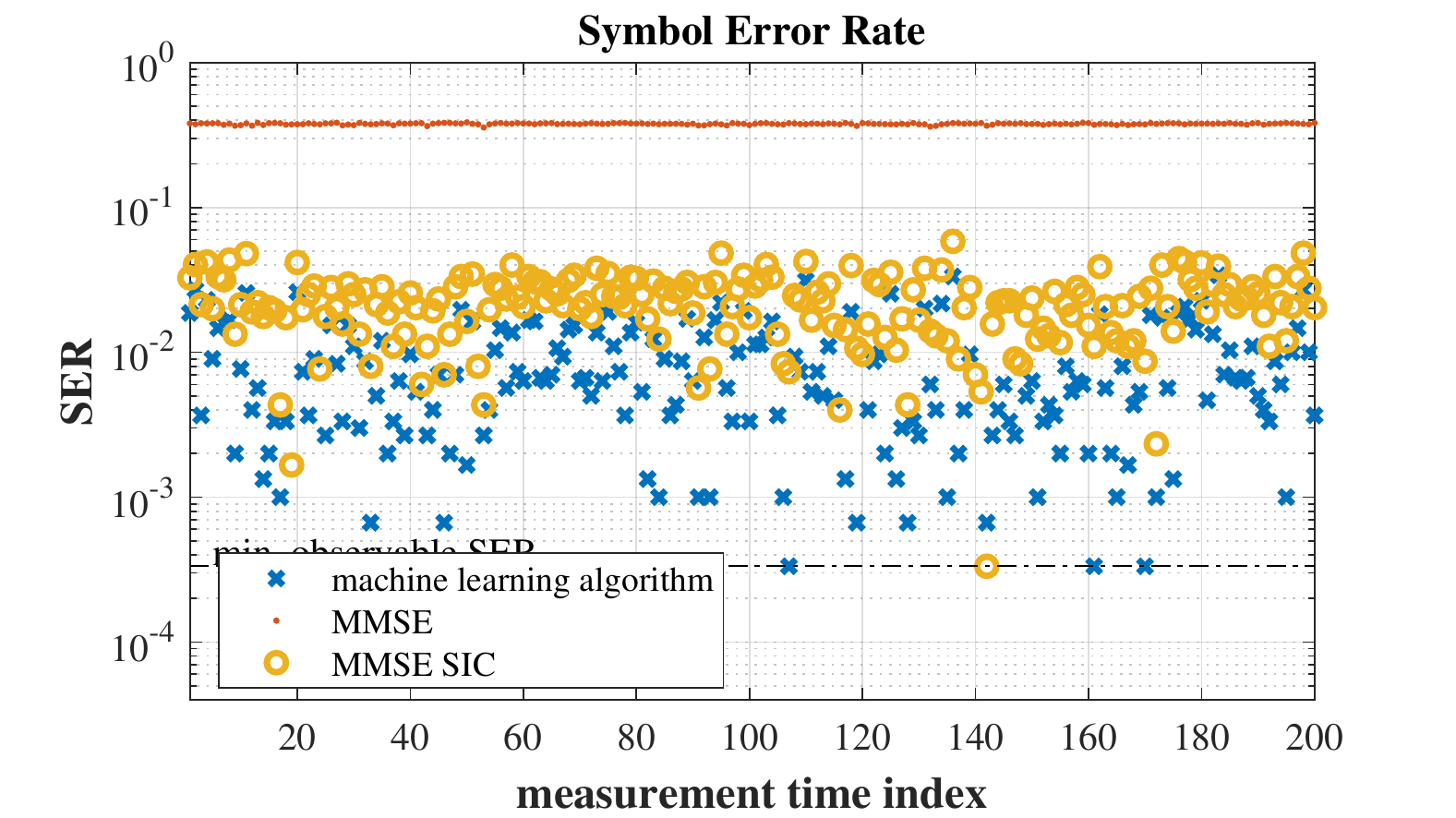}
    \caption[SER6]{User 6: \(  \omega_{L} = 0.6 \) and \( \omega_{G} = 0.4 \) with \( T_{length} = 685 \)}
    \label{fig:SER6}
   \vspace*{-1.0\floatsep}
\end{figure}

\begin{figure}[b!]
   \vspace*{-1.0\floatsep}
    \captionsetup{skip=0.2\baselineskip,size=footnotesize}
    \centering
    \includegraphics[width=0.97\linewidth,clip=true]{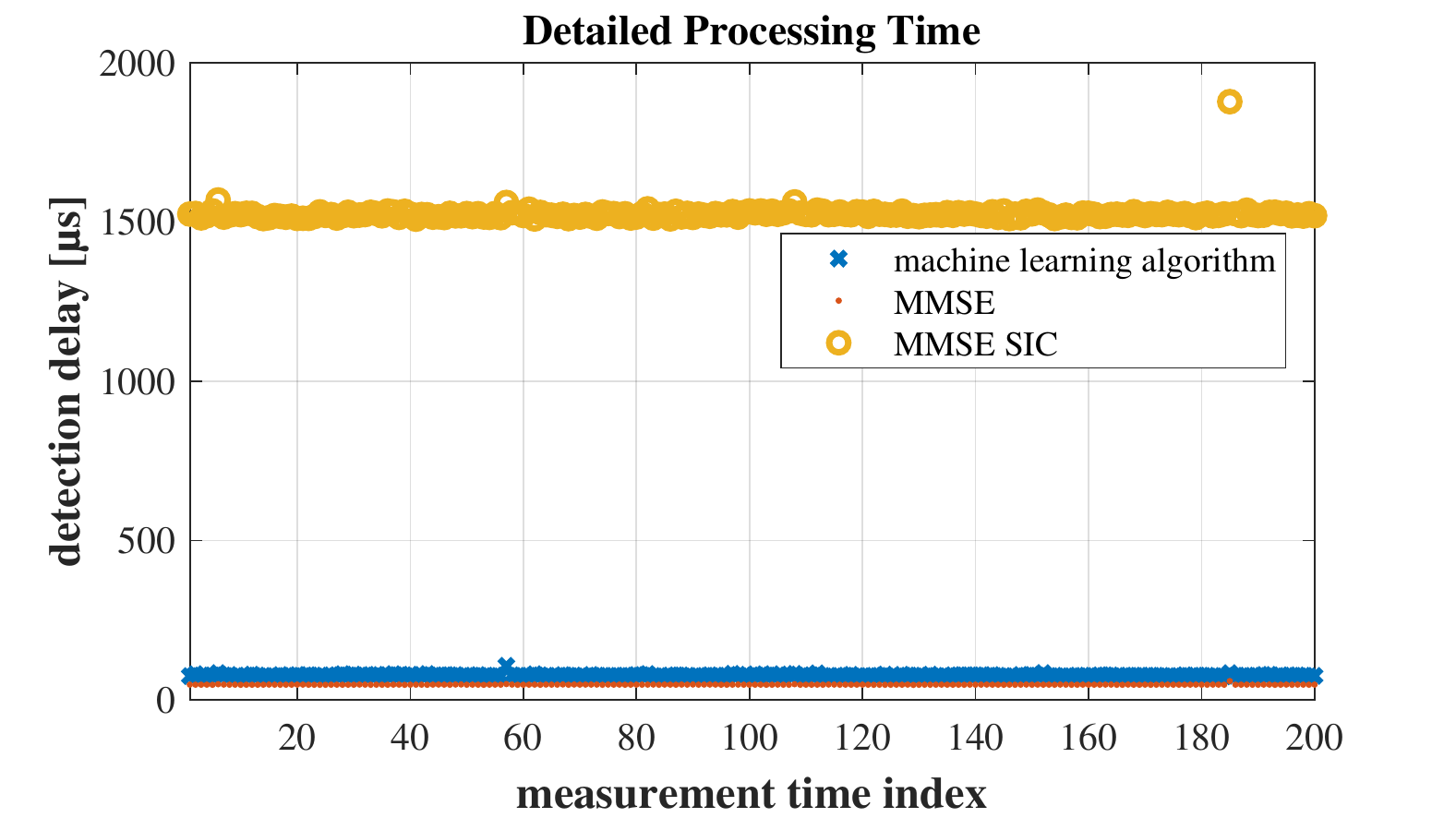}
    \caption[DELAY5]{User 5: \(  \omega_{L} = 0.6 \) and \( \omega_{G} = 0.4 \) with \( T_{length} = 685 \)}
    \label{fig:DELAY5}
    \vspace*{+0.5\floatsep}
    %
%
    \captionsetup{skip=0.2\baselineskip,size=footnotesize}
    \centering
    \includegraphics[width=0.97\linewidth,clip=true]{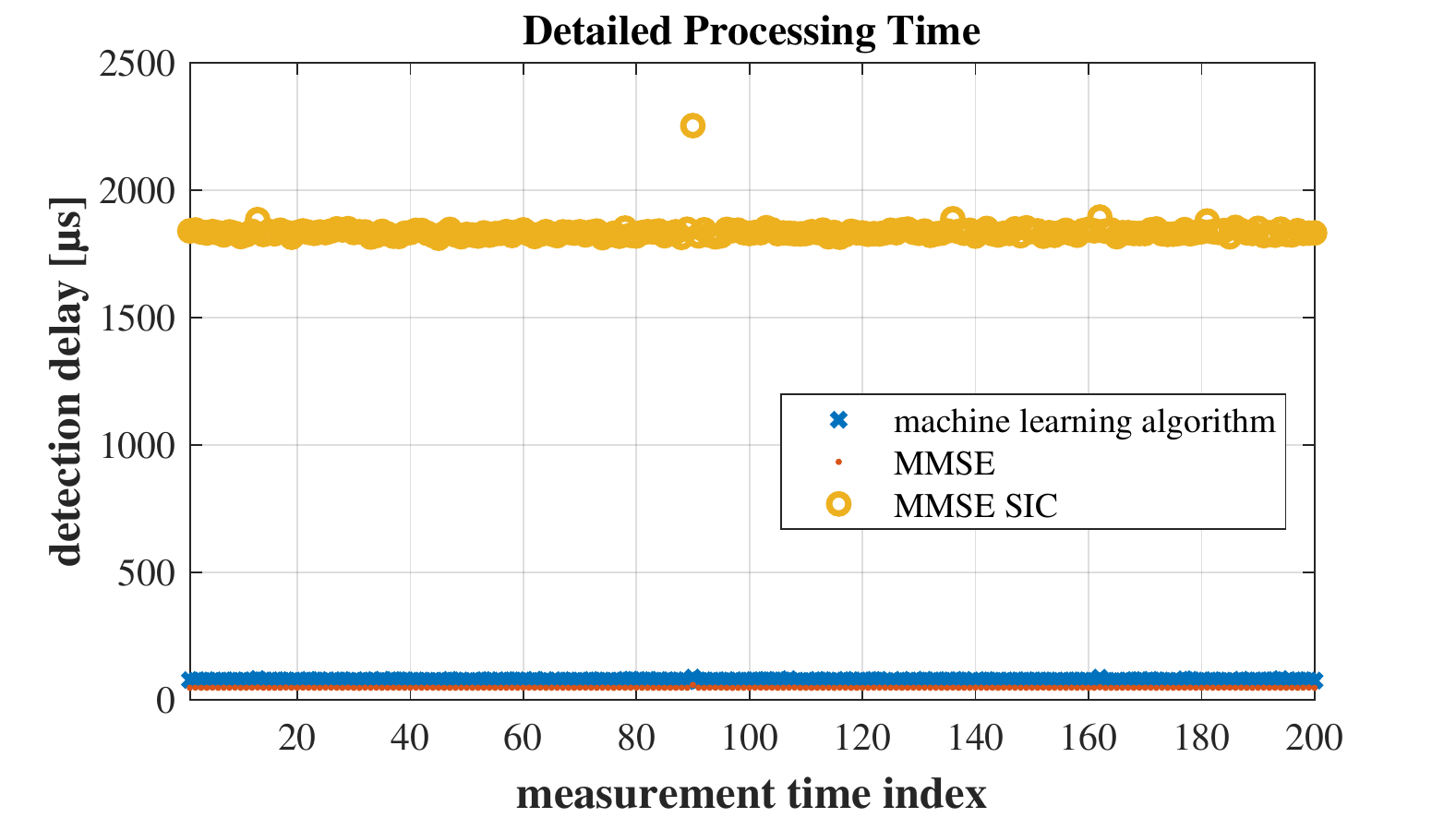}
    \caption[DELAY6]{User 6: \(  \omega_{L} = 0.6 \) and \( \omega_{G} = 0.4 \) with \( T_{length} = 685 \)}
    \label{fig:DELAY6}
    \vspace*{+0.5\floatsep}
    %
%
    \captionsetup{skip=0.2\baselineskip,size=footnotesize}
    \centering
    \includegraphics[width=0.97\linewidth,clip=true]{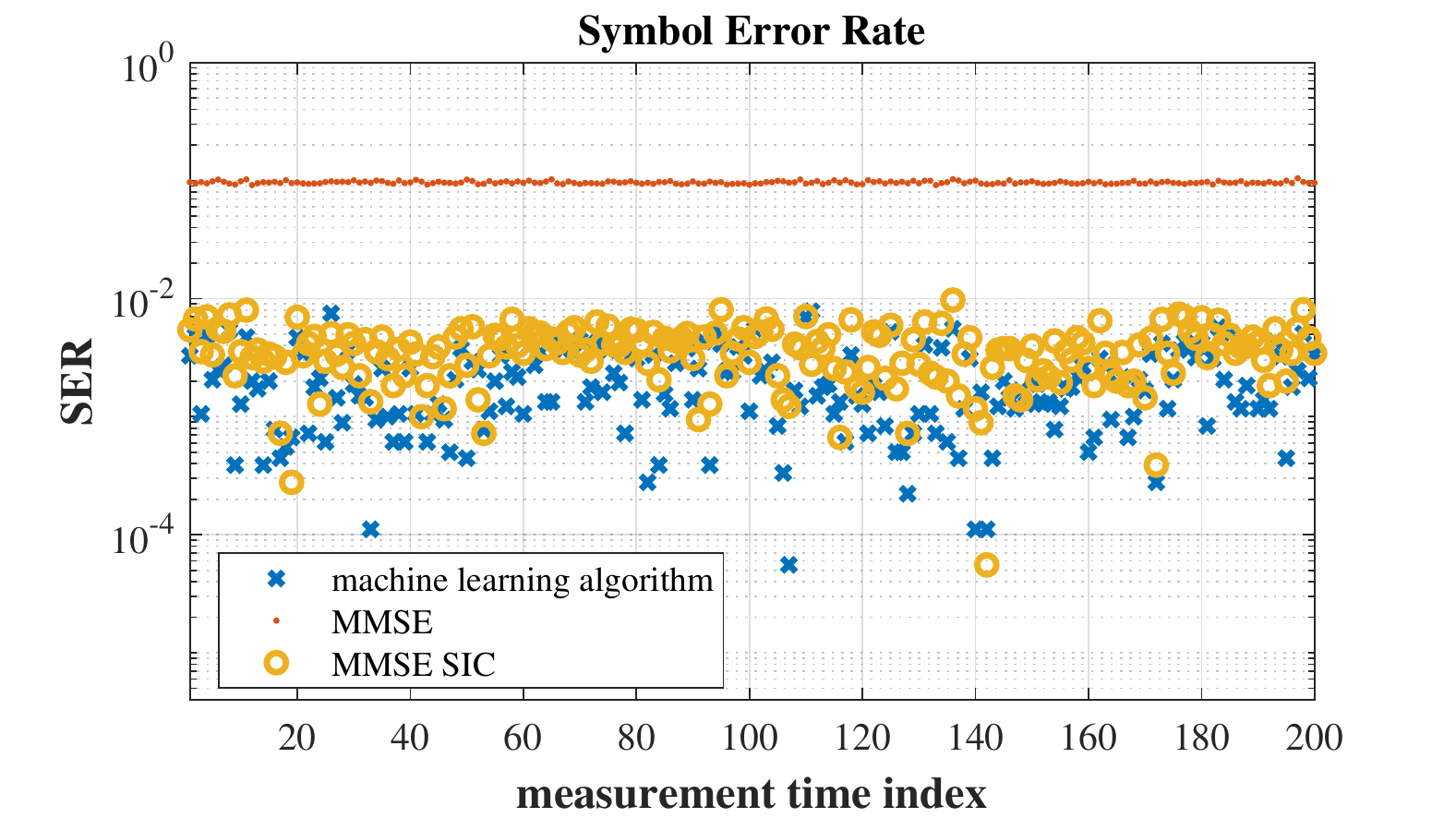}
    \caption[SERall]{Mean over all users: \(  \omega_{L} = 0.6 \) and \( \omega_{G} = 0.4 \)}
    \label{fig:SERall}
    \vspace*{+0.5\floatsep}
    %
%
    \captionsetup{skip=0.2\baselineskip,size=footnotesize}
    \centering
    \includegraphics[width=0.97\linewidth,clip=true]{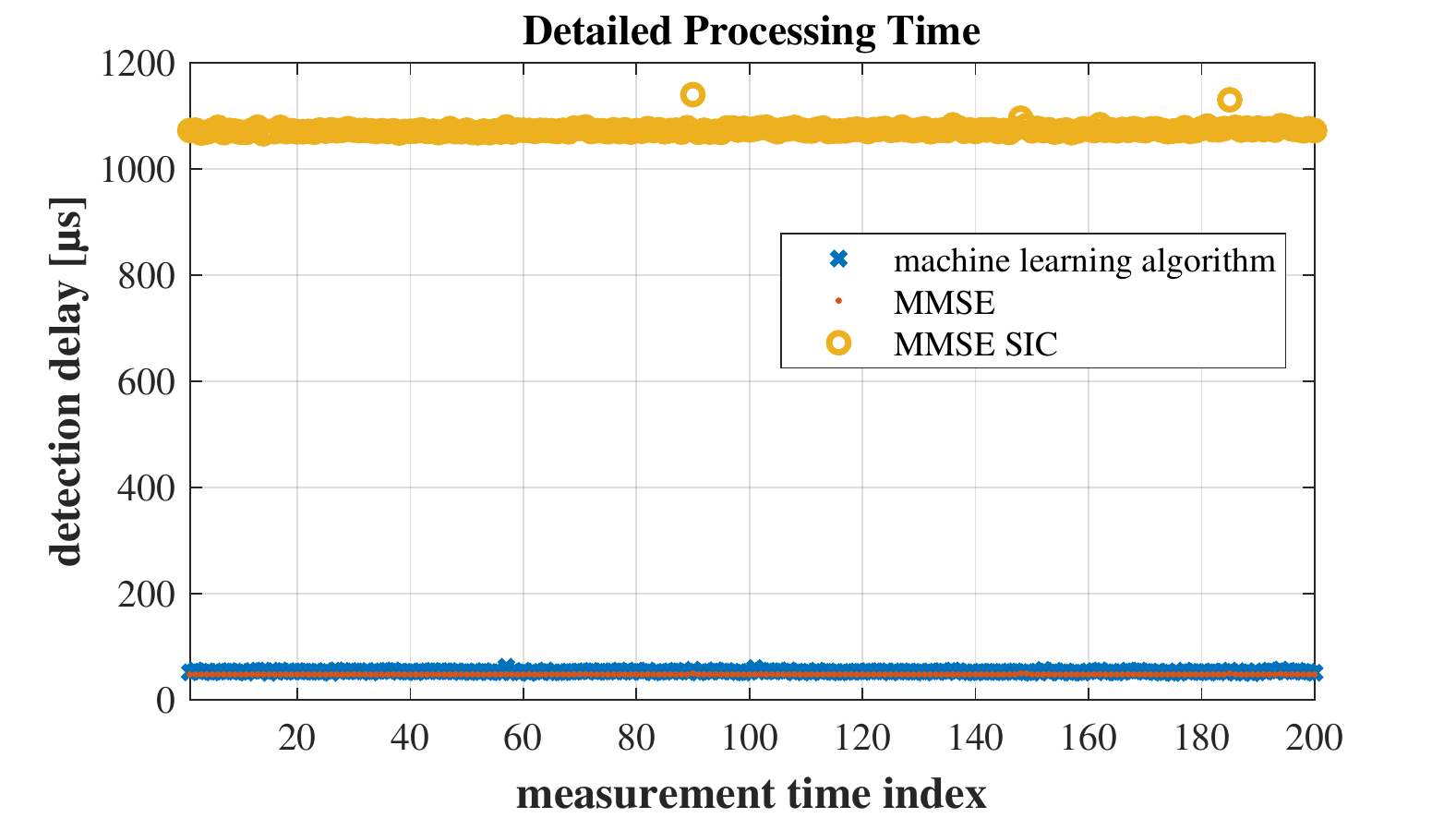}
    \caption[DELAYall]{Mean over all users: \(  \omega_{L} = 0.6 \) and \( \omega_{G} = 0.4 \)}
    \label{fig:DELAYall}
    %
    %
\end{figure}

\Cref{fig:SER5} and \Cref{fig:SER6}, show the SER results for the $5$th and the $6$th user, respectively. We see that the \ac{MMSE} shows a poor performance in both cases because these users are linearly inseparable and they suffer from excessive interference from other users. In contrast, the \ac{MMSE}-\ac{SIC} and \ac{NL}-\ac{ML} show a good and comparable performance in this case. \Cref{fig:SERall} shows the average SER of the system. We see that both \ac{MMSE}-\ac{SIC} and \ac{NL}-\ac{ML} show a good performance, where \ac{NL}-\ac{ML} slightly outperforms \ac{MMSE}-\ac{SIC}. 

\Cref{fig:DELAY5} and \Cref{fig:DELAY6}, show the complexity (measured in terms of detection delay) results for the $5$th and the $6$th user, respectively. We see that the \ac{MMSE} has low complexity because it is a linear technique. Even though \ac{NL}-\ac{ML} is a nonlinear technique, one of its strengths is that it involves robust linear processing and all users are detected in parallel independently. We see that the complexity of \ac{NL}-\ac{ML} is comparable to that of the \ac{MMSE} technique. As expected, the nonlinear \ac{MMSE}-\ac{SIC} has a higher complexity than both \ac{MMSE} and \ac{NL}-\ac{ML}. \Cref{fig:DELAYall} shows the average complexity of the system where we observe once again that \ac{NL}-\ac{ML} receiver has a low complexity. 

 In conclusion, the simulation results demonstrate that the \ac{NL}-\ac{ML} receiver is capable of achieving the SER performance of nonlinear techniques, while it has a complexity comparable to that of linear techniques.  
 
%% file: sections/conclusions.tex
\section{Conclusions}
\label{sec:Con}

We demonstrated a nonlinear \ac{ML} based multiuser receiver that can achieve the performance of nonlinear receivers and it has the complexity comparable to linear receivers.
The proposed receiver does not require parameter estimation (user channels, powers, noise variance, etc.) as in conventional receivers which is subject to errors.
We compared the performance with the standard linear \ac{MMSE} receiver and also with the nonlinear \ac{MMSE}-\ac{SIC} in a hardware-in-the-loop system. The results show that the proposed \ac{ML} based multiuser is simpler yet powerful alternative to conventional nonlinear receivers. 

\comment{
\marginnote{\textcolor{red}{Summary}}
In this work, we show an experimental setup to examine hardware impairments of an \ac{NOMA} enabled \ac{SDR} system.
For this demonstration, a commercial available flexible \ac{SDR} platform called NAMC-SDR with eight antenna ports in receive as well as transmit direction is used.
We compare the result of our proposed algorithm with \ac{MMSE} and \ac{MMSE}-\ac{SIC} for six users in a cluster.
Finally, the proposed demonstration system shows the expected behavior we know from the simulations in \cite{8422449}.

\marginnote{\textcolor{red}{What  do the results mean}}
...
\vspace*{0.5cm}

\marginnote{\textcolor{red}{How do they relate to other published results?}}
We build a live demonstrator based on the  online adaptive \ac{ML} approach shown in \cite{8422449}.
The important related papers regarding signal processing in multi kernel \ac{RKHS} are \cite{5256321} and \cite{7174566}.

\marginnote{\textcolor{red}{What are the implications?}}
The practical nonlinear \ac{ML} based receive filtering technique enables a simple and reliable multi user detection in a wireless communication system.
This method has low complexity and works with short training and a small number of receive antennas; nevertheless, all users are detected in parallel independently.
The demo shows that \ac{ML} can replace some building blocks of wireless receivers and that the method outperforms other conventional non-linear schemes. 

\marginnote{\textcolor{red}{What problems occurred?}}
From the hardware perspective view we decided to use an oversampled raised cosine filtered \ac{BPSK} symbol to keep the baseband bandwidth small and therefor extending the channel coherence time.
But for a high number of samples, a rotation becomes visible introduced by a small difference between transmitter and receiver \acp{LO}, even when they are using the same reference clock or \ac{CFO} is measured and compensated.
We avoid any time synchronization issues between all active users by using the same \ac{SDR}.
From the software perspective view finding the perfect parameter setting for each user is still an open question.

\marginnote{\textcolor{red}{What improvments could be made}}
We also don't address the cell-less aspect in this paper.
This aspect is an extension of the proposed scheme such that the \ac{UL} signals will be received by multiple \acp{gNB} and combined in a central processing unit, considering limited backhaul.

\marginnote{\textcolor{red}{What more needs to be done}}
Instead of mapping the \ac{BPSK} symbol to a smalband \ac{SCM} in time domain we would like to map to a \ac{MCM} like \ac{OFDM} or \ac{SC-FDMA} in frequency domain, which is more common in current mobile communication systems.

}

%% file: sections/acknowledgments.tex

\section*{Acknowledgments}
\addcontentsline{toc}{section}{Acknowledgments}

The work is supported by the \acl{EC} and \ac{5GPPP} and received funding from the \acs{EC} \acs{H2020}/\acs{5GPPP} program ONE5G (ICT-760809) project\cite{ONE5G}.
